\documentclass[sigconf,anonymous=false,balance=true]{acmart}
\AtBeginDocument{%
  }

\copyrightyear{2026}
\acmYear{2026}
\setcopyright{othergov}
\acmConference[SIGIR '26]{Proceedings of the 49th International ACM SIGIR Conference on Research and Development in Information Retrieval}{July 20--24, 2026}{Melbourne, VIC, Australia}
\acmBooktitle{Proceedings of the 49th International ACM SIGIR Conference on Research and Development in Information Retrieval (SIGIR '26), July 20--24, 2026, Melbourne, VIC, Australia}
\acmDOI{10.1145/3805712.3809565}
\acmISBN{979-8-4007-2599-9/2026/07}

\makeatletter
\def\@ACM@copyright@check@cc{}
\makeatother

\usepackage{xcolor}
\usepackage{colortbl}
\usepackage{pdfpages}
\usepackage{array}
\usepackage{multirow}
\usepackage{makecell}
\usepackage{paralist}
\usepackage{enumitem}
\usepackage{graphicx}
\usepackage{bbm}
\usepackage{pifont}

\begin{document}

\title{From Tokens to Concepts: Leveraging SAE for SPLADE}
\author{Yuxuan Zong}
\orcid{0009-0002-0376-1369}
\affiliation{%
  \institution{Sorbonne Université, CNRS, ISIR}
  \city{Paris}
  \country{France}
}
\email{yuxuan.zong[at]isir.upmc.fr}

\author{Mathias Vast}
\orcid{0009-0007-4612-717X}
\affiliation{%
  \institution{Sinequa by ChapsVision}%
  \institution{Sorbonne Université, CNRS, ISIR}
  \city{Paris}
  \country{France}
}
\email{vast[at]isir.upmc.fr}

\author{Basile Van Cooten}
\orcid{0009-0002-0234-917X}
\affiliation{%
  \institution{Sinequa by ChapsVision}
  \city{Paris}
  \country{France}
}
\email{bvancooten[at]chapsvision.com}

\author{Laure Soulier}
\orcid{0000-0001-9827-7400}
\affiliation{%
  \institution{Sorbonne Université, CNRS, ISIR}
  \city{Paris}
  \country{France}
}
\email{laure.soulier@isir.upmc.fr}

\author{Benjamin Piwowarski}
\orcid{0000-0001-6792-3262}
\affiliation{%
  \institution{CNRS, Sorbonne Université, ISIR}
  \city{Paris}
  \country{France}
}
\email{benjamin.piwowarski@cnrs.fr}

\renewcommand{\shortauthors}{Zong et al.}

\newcommand{\mcL}{\mathcal{L}}
\newcommand{\bw}{\mathbf{w}}
\newcommand{\bW}{\mathbf{W}}
\newcommand{\bb}{\mathbf{b}}
\newcommand{\bh}{\mathbf{h}}
\newcommand{\be}{\mathbf{e}}
\newcommand{\RR}{\mathbb{R}}
\newcommand{\bE}{\mathbf{E}}
\newcommand{\bQ}{\mathbf{Q}}
\newcommand{\bD}{\mathbf{D}}
\newcommand{\bH}{\mathbf{H}}
\newcommand{\bC}{\mathbf{C}}
\newcommand{\bone}{\mathbbm{1}}
\newcommand{\cmark}{\ding{51}}%
\newcommand{\xmark}{\ding{55}}%

\newcommand{\todo}[1]{\textcolor{red}{[TODO #1]}}

\newcommand{\EEscore}{\mathbf{E}^2}
\newcommand{\mrr}{\operatorname{MRR}}
\newcommand{\flops}{\operatorname{QD-FLOPs}}
\newcommand{\softplus}{\operatorname{softplus}}

\newcolumntype{L}[1]{>{\raggedright\arraybackslash}p{#1}}
\newcolumntype{C}[1]{>{\centering\arraybackslash}p{#1}}
\newcolumntype{R}[1]{>{\raggedleft\arraybackslash}p{#1}}

\begin{abstract}

  Learned Sparse IR models, such as SPLADE, offer an excellent efficiency-effectiveness tradeoff. However, they rely on the underlying backbone vocabulary, which might hinder performance (polysemicity and synonymy) and pose a challenge for multi-lingual and multi-modal usages.
  To solve this limitation, we propose to replace the backbone vocabulary with a latent space of semantic concepts learned using Sparse Auto-Encoders (SAE). 
  Throughout this paper, we study the compatibility of these 2 concepts, explore training approaches, and analyze the differences between our SAE-SPLADE model and traditional SPLADE models. 
  Our experiments demonstrate that SAE-SPLADE achieves retrieval performance comparable to SPLADE on both in-domain and out-of-domain tasks while offering improved efficiency.
  
\end{abstract}



\maketitle

\section{Introduction}


Pre-trained contextualized Language Models (PLM) based on the Transformer architecture, such as BERT~\cite{devlin2019bert}, significantly improved retrieval effectiveness over previous state-of-the-art methods in Information Retrieval (IR) such as BM25~\cite{Robertson1994OkapiAT}. Following their introduction, a wide range of methods appeared. Dual encoders (either sparse~\cite{formal2021spladev2, gao2021coil} or dense~\cite{karpukhin2020dense}), as well as late-interaction models, such as ColBERT~\cite{santhanam2022colbertv2}, are considered the most efficient and effective and are traditionally used as first-stage retrievers.

Among first-stage retrievers, Learned Sparse Retrieval (LSR) models like SPLADE \citep{formal2021splade, formal2021spladev2} have shown great potential in the era of PLM. 
Unlike traditional lexical models~\citep{Robertson1994OkapiAT}, LSR models are able to alleviate the problem of lexical mismatch by learning better term weights and performing term expansion based on contextualized embeddings. 
LSR models further retain (to some extent) the good properties of Bag-of-Words models, such as the exact matching of terms or interpretability, and benefit from the proven efficiency of inverted indexes, which allows for fast retrieval. 

\begin{figure}
    \centering
    \includegraphics[width=\columnwidth]{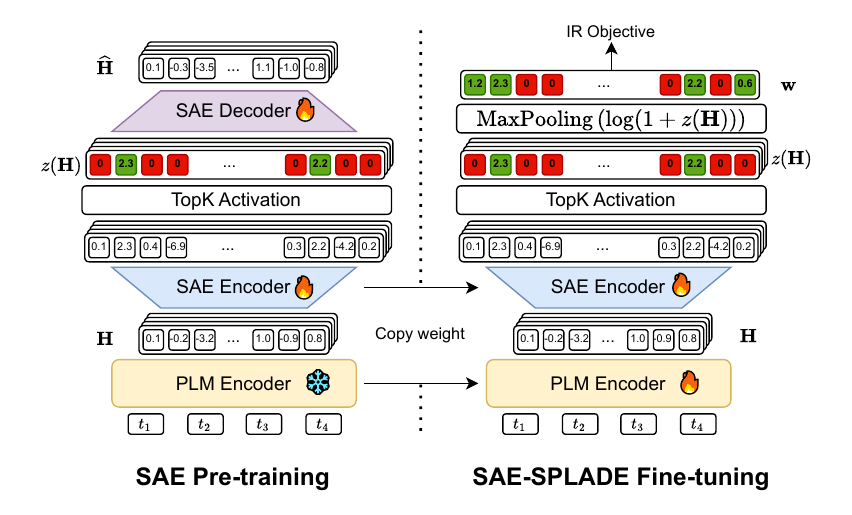}
    \caption{The architecture of our SAE-SPLADE model. We start with SAE pre-training (left). The SAE-SPLADE is obtained by removing the SAE decoder and adding the SPLADE document-level aggregation (right).} 
    \label{fig:archi}
\end{figure}

However, most well-performing LSR models still rely on the encoder vocabulary. The reason stems from the MLM pre-training that pre-conditions the model for learning sparse neural IR representations.
This has several implications.
First, adapting to new languages (or modalities) is not straightforward~\cite{chen2024bge,nguyen2025milco,lassance2023extending}. 
Second, the model relies on term expansion to address \emph{vocabulary mismatch} (e.g., synonyms) and splits entity mentions into parts~\cite{nguyen2024dyvo}, which is not fully satisfactory.
Third, the model derives the sparse representation of a text by using the hidden states of the last layer. However, the final Transformer layer is not necessarily the most suitable for many NLP/IR tasks~\citep{skean2025layer}.


Recently, Sparse Auto-Encoders (SAEs) have been introduced to uncover concepts manipulated by Large Language Models (LLM) by projecting latent states into a (sparse) vector of SAE concepts~\citep{gao2025scaling,rajamanoharan2024jumping}. 
In the context of LSR models, this is interesting since it allows replacing the MLM projection layer with an SAE one.
Recent works~\cite{lassance2021composite, park2025decoding, kang2025interpret, lassance2021composite, wen2025beyond, formal2026learning} have shown that projecting the document/query representation into the SAE concept space is already promising.
%

In this work, we go further and study in depth how SAE can be used to replace the MLM head of SPLADE and be applied to the IR task, with a new SAE-SPLADE architecture.
More generally, we explore the following research questions:
\begin{description}
    \item[RQ1]  \emph{Is it possible to replace the token output space of LSR models with a latent output space using SAE? }
    \item[RQ2] \emph{What benefits does SAE bring to LSR models in terms of efficiency and effectiveness? }
    \item[RQ3] \emph{Is SAE a suitable solution to the limitations of SPLADE with multilingual retrieval? }
    \item[RQ4] \emph{What are the differences between SAE-SPLADE and SPLADE? }
\end{description}

In this paper, we show how to combine SAE and SPLADE into a new SAE-SPLADE model evaluated across a range of In-Domain (ID) and Out-of-Domain (OoD) datasets. Our results show that SAE-SPLADE models are able to achieve state-of-the-art performance while being more efficient than their SPLADE counterparts.  To compare different effectiveness-efficiency tradeoffs between models, we introduce a new scoring function $\EEscore$ (for Efficiency-Effectiveness), which could serve as a basis for future work in sparse neural IR. We also conduct an extensive series of experiments to study the impact of key hyperparameters and design choices on our proposed method. Finally, we explore the application of SAE-SPLADE to multilingual retrieval, showing that its semantic vocabulary can drastically increase document sparsity compared to traditional SPLADE~\citep{formal2021spladev2}, and present a qualitative analysis to illustrate the differences in behavior between our proposed SAE-SPLADE and SPLADE. The code used for the experiments in this paper can be accessed online\footnote{\href{https://github.com/yzong12138/sae_splade}{https://github.com/yzong12138/sae\_splade}}

\section{Related Works and Preliminaries}


LSR refers to a family of first-stage retrieval methods that are trained to predict sparse vector representations of queries and documents, enabling efficient retrieval. 
Before the Transformer era, \citeauthor{zamani2018from} proposed SNRM, which projects text representations in a sparse space for efficient retrieval. SNRM relies on an initial random projection, which limits its effectiveness in practice. Subsequent work~\cite{macavaney2020expansion,formal2021splade,formal2021spladev2, nguyen2023unified}, including SPLADE, replaced SNRM with the Transformer architecture and leveraged the MLM projection head, which preconditions the model for IR~\cite{FormalEffectiveEfficientSparse2023}. This led to state-of-the-art first stage rankers~\cite{lassance2024spladev3} when combined with effective training strategies such as distillation~\cite{formal2022distillation}.

As argued in the introduction, we hypothesize that these models still suffer from several limitations linked to the reliance on the PLM vocabulary.
%
Subsequently, some works have explored ways to untie the input and output vocabularies~\cite{dudek2023learning,kim2025role,nguyen2024dyvo, yu2024improved}. However, these works still rely on the original vocabulary projection matrix to compute the new output vocabulary, as the construction of sparse retrieval models based on randomly initialized vocabularies has been shown to be detrimental to performance~\cite{zamani2018from}. 
In contrast, we propose using sparse autoencoders to learn a \emph{semantic vocabulary}, which substantially alleviates these limitations.

%


\label{sec:related_sae}

Orthogonally, the success of LLMs~\cite{touvron2023llama, yang2025qwen3, brown2020language} has led to a deeper investigation of their behavior.
Among these interpretability works, SAEs~\cite{ng2011sparse} are particularly interesting in the context of LSR: SAEs disentangle the complex semantic structures intertwined in the dense embeddings of these models into a set of distinct and interpretable conceptual units. 

In practice, an SAE encodes a hidden state $\bh \in \RR^d$ in a sparse vector of latents $z(\bh)\in \RR^{M}$. Its encoder consists of a linear transformation $\bW_{enc}\in \RR^{M\times d}$, with bias $\bb_{enc} \in \RR^M$, followed by an activation function $\phi$, e.g. $\operatorname{ReLU}$: 
\begin{equation}
    z(\bh) = \phi(\bW_{enc} \bh  + \bb_{enc})
\end{equation} 
The overall number of latents $M$ is usually chosen to be much larger than the original hidden dimension size ($M \gg d$) to allow the encoder to learn a more expressive representation of the original input (overcomplete basis). 

To train this encoder, SAEs models rely on a \emph{decoder} that predicts $\widehat \bh$ from $z(\bh)$, keeping $\bh$ and $\widehat \bh$ as close as possible. 
Formally, we have 
\begin{equation}
    \begin{split}
        & \widehat{\bh} = \operatorname{SAE}(\bh) = \bW_{dec} z(\bh)  + \bb_{dec} \\
        & \mcL_{rsct} = \|\widehat{\bh} - \bh\|^2_2
    \end{split}
\end{equation}
where $\bW_{dec}\in\RR^{d\times M}$ and $\bb_{dec}\in \RR^{d}$ are the decoder weight matrix and bias term.
To ensure that the representation remains sparse, a regularization loss $\mcL_{sparsity}$, which typically minimizes the $\ell_1$ norm of the sparse activation vector, is used. Without this sparsity constraint, the SAE could maintain a high reconstruction quality by keeping a large number of latents activated (e.g., $d$).
The overall loss, $\mcL_{SAE}$,  can be defined as: 
\begin{equation}
    \mcL_{SAE}(\bh) = \mcL_{rsct} + \alpha_{sp} \mcL_{sparsity}
    \label{eq:sae_loss}
\end{equation}
where $\alpha_{sp}$ controls the importance of the sparsity regularization.

Since its inception, SAE~\cite{ng2011sparse} has been improved -- a non-exhaustive list includes: Gated SAE \citep{rajamanoharan2024improving}, TopK SAE \citep{gao2025scaling}, BatchTopK SAE \citep{bussmann2024batchtopk} or JumpRelu SAE \citep{rajamanoharan2024jumping}. All aim at solving two known issues of the original method: (1) the \textit{shrinkage bias problem}, where the $\ell_1$ norm designed to enforce the sparsity also contributes to minimizing the magnitude of the latents activation, reducing reconstruction accuracy; and 2) the \textit{``dead'' latents issue}, where certain latents cease to activate entirely during training (as their activation values is always $0$). 
Among those, TopK SAE~\citep{gao2025scaling} leverages a top-$k$ filtering on each embedding $z(\bh)$, avoiding to explicitly control the sparsity through regularization. 
Other works developed TopK SAE variants that improve the structure of the latent space. For example, Hierarchical TopK SAE \cite{balagansky2025train} learn a single model while optimizing reconstructions at multiple sparsity levels, while Matryoshka TopK SAE~\cite{bussmann2025learning} train simultaneously multiple nested SAE models of increasing sizes. 
This training objective leads coarse-grained semantics to be captured by the first $k$ latents, while finer-grained information is progressively encoded in later dimensions.

In IR, SAEs have been used to approximate dense representations for efficient nearest-neighbor search~\citep{park2025decoding,kang2025interpret}.
In particular, \citeauthor{park2025decoding}~\citep{park2025decoding} show that SAE-derived features are effective indexing units. However, all prior approaches combine SAE with already-trained dense retrievers.
More recently, SPLARE~\citep{formal2026learning} trains an LSR using SAE latents derived from an LLM~\citep{lieberum2024gemma}, under a SPLADE objective. However, the SAE component is frozen, limiting the model’s ability to adapt to different feature spaces for specific tasks. Moreover, its effectiveness on smaller backbones, commonly used in earlier work~\cite{formal2021spladev2}, remains unclear. 
In contrast, we propose to train an SAE directly from a PLM backbone on a general IR corpus, i.e., MS MARCO \citep{bajaj2016msmarco}, before combining it with a SPLADE model. The resulting SAE-SPLADE is then fine-tuned on the IR task, using the SAE module instead of its original MLM head, enabling a tighter integration between relevance modeling and sparsity during training.

\section{Connecting SAE with SPLADE}

%
As discussed above, the multiple shortcomings of LSR models stemming from their MLM head and its tight dependence on the input vocabulary lead us to introduce a new architecture, no longer bound by these constraints.
Although promising, SNRM has shown that relying on a random projection head (i.e. random vocabulary vectors) does not pre-condition the model sufficiently for state-of-the-art performance~\cite{zamani2018from} -- and our experiments confirm it.
Therefore, in contrast to previous work, we introduce a completely new semantic vocabulary learned with SAE. Each element of this vocabulary—each latent—represents a specific semantic concept rather than a fixed token. We conjecture that this design mitigates both vocabulary and semantic mismatches, enabling the model to transfer more effectively across contexts, including to unseen languages. Specifically, unlike the original SPLADE model, our SAE-SPLADE and its latent-based vocabulary are expected to generate representations that are more consistent across diverse languages and/or modalities, thereby enhancing transferability in multilingual retrieval settings.

As illustrated in \autoref{fig:archi}, to combine SAE with a sparse neural IR model, we use a two-stage training pipeline. First, we train an SAE by reconstructing the contextualized token embeddings. We then use only its encoder to project the contextualized embeddings into a (sparse) latent space for relevance estimation.  The trained SAE encoder $z(\cdot)$, which outputs sparse representations in $\mathbb{R}^M$, projects the contextualized representations $\bH = (\bh_0, \bh_1, \dots, \bh_N)$ of the PLM, similar to what SPLADE does with the original MLM projection head~\citep{formal2021splade}. We then use the same activation function and pooling mechanisms as SPLADE-max~\citep{formal2021spladev2}.



\paragraph{SAE Training}

Training an SAE relies on dense contextualized embeddings produced by a pre-trained model. In this work, we use encoder-based architectures such as BERT~\citep{devlin2019bert} or DistilBERT~\citep{sanh2019distilbert} to encode token-level embeddings from the corpus, before feeding them to the SAE. Importantly, the PLM encoder's weights are kept frozen when training the SAE.
Among the various SAE versions cited above, we first experimented with JumpReLU SAE~\citep{rajamanoharan2024jumping} but it proved too unstable\footnote{Official SAELens documentation reports this issue: \url{https://github.com/decoderesearch/SAELens/blob/main/docs/training_saes.md}},
and used the widely used TopK SAE~\citep{gao2025scaling}. We also experimented with its variants, Hierarchical TopK SAE~\citep{balagansky2025train} and Matryoshka TopK SAE~\citep{bussmann2025learning}, in section \ref{sec:k_effect}.


\paragraph{Sparse Retrieval Training}

To fine-tune SAE-SPLADE, we substitute the encoder’s original MLM head with the trained SAE encoder and drop the SAE decoder. In contrast to the previous stage, where only the SAE is trained, we now also fine-tune the original encoder parameters.
%
%
We use the same fine-tuning objective as SPLADEv3~\cite{lassance2024spladev3}, which combines both the KL-divergence ($\mathcal{L}_{KL}$) and MarginMSE ($\mathcal{L}_{MSE}$) distillation objectives, but not their complex data generation strategy -- e.g., self-distillation~\cite{formal2022distillation}, a stronger cross-encoder to score negatives, and the curriculum learning approach~\cite{zeng2022curriculum}.
%

Although the top-$k$ constraint of TopK SAE already controls sparsity, our empirical study shows that the overall sparsity over the aggregated document (or query) representation is still relatively low. Following SPLADE~\citep{formal2021splade,formal2021spladev2}, we also add the FLOPs regularization $\mathcal{L}_{flops-d}$ and $\mathcal{L}_{flops-q}$~\citep{paria2020minimizing}, to further control the sparsity of SAE-SPLADE representation when learning the IR objective. 
%
Overall, our training objective for SAE-SPLADE is:
\begin{equation}
    \begin{split}
    \mcL_{\operatorname{SAE-SPLADE}} & = \lambda_{\operatorname{KL}} \mcL_{\operatorname{KL}} + \lambda_{\operatorname{MSE}} \mcL_{\operatorname{MSE}} \\
    & + \lambda_{\operatorname{flops-d}} \mcL_{\operatorname{flops-d}} + \lambda_{\operatorname{flops-q}} \mcL_{\operatorname{flops-q}}
    \end{split}
\end{equation}

Note that we also experimented jointly training SAE and SPLADE, but found that it was detrimental. We attribute this to the misalignment between the reconstruction loss and the IR task objective, as one tries to compress as much as possible information that is essential to the other. However, as shown later, SAE (as MLM) provides a good pre-conditioning for the training of SPLADE.

\section{Experiments and Results}


\subsection{Experimental Setup}

\paragraph{Dataset} 

We train the SAE on the 8.8M passages of the MS MARCO v1 dataset~\cite{bajaj2016msmarco} and fine-tune SAE-SPLADE with the same dataset as ColBERTv2~\citep{santhanam2022colbertv2} (distillation based on a cross-encoder). We evaluate the models, \emph{in-domain}, on the MS MARCO dev small set (6980 queries), as well as on the 2019 (43 assessed topics) and 2020 (54 assessed topics) sets of the TREC-DL track \cite{craswell2019overview, craswell2020overview} (Avg. of 2 sets). We report OoD performance on LoTTE Search~\citep{santhanam2022colbertv2} (average of 5 subsets) for all preliminary and ablation experiments, and additionally compare SAE-SPLADE to the other baselines on the 13 publicly available datasets of BEIR, following the common practice. 

\paragraph{Implementation Details} \label{sec:train_config}

We conduct all our experiments with a single A100 GPU\footnote{PyTorch 2.8.1,  HuggingFace transformers 4.37}. We initialize our model with DistilBERT~\citep{sanh2019distilbert} (pre-trained checkpoint from HuggingFace\footnote{\texttt{distilbert/distilbert-base-uncased}}), since it is much smaller than a BERT model (2x speed gain) while being competitive, as acknowledged in SPLADEv2~\citep{formal2021spladev2, formal2022distillation}. During training, we use the AdamW optimizer with a learning rate of 5e-5, using 768 documents per batch ($\approx$60k tokens) for 160k steps. 


To initialize the SAE parameters, we follow previous work~\citep{gao2025scaling,rajamanoharan2024jumping} and set the biases $\bb_{enc}$ and $\bb_{dec}$ to zero and the encoder projection matrix to the transposed of the decoder projection matrix ($\bW_{enc} = \bW_{dec}^\top$), but do not tie them during SAE training. Throughout training, we re-normalize the latent vectors to the unit norm after each step. 


We fine-tune SAE-SPLADE for 240k steps, with a learning rate of 2e-5, $\lambda_{\operatorname{KL}}=1$, $\lambda_{\operatorname{MSE}}=0.05$, $\lambda_{\operatorname{flops-d}}=0.04$ and $\lambda_{\operatorname{flops-d}}=0.06$. Each training batch is composed of 32 queries, with 8 hard negatives for each.
For both training stages and the evaluation, we truncate queries to 32 tokens and documents to 256 tokens.\footnote{Except for the ArguAna dataset in BEIR, where queries were also truncated to 256 tokens, similar to ColBERT and SPLADE, as the queries themselves long documents.}
We use the Triton kernel provided by~\citet{gao2025scaling} to maintain a comparable training and indexing speed, even when increasing the size of the latent space. With this setup, SAE training takes around 35 hours and IR optimization around 24 hours ($\approx$ 2.5 days to train SAE-SPLADE from scratch).

\subsection{Evaluation of Sparse Models} \label{sec:eval_sparse}

Besides the standard IR metrics (e.g., MRR@10 or nDCG@10) used to quantify the effectiveness of the models, we also report their efficiency using the average \emph{Query-Document FLOPs} (or QD-FLOPs) between the query and document representations. This metric estimates the expected number of posting list entries, per query and document, that need to be accessed during retrieval (with a naive retrieval approach). Formally\footnote{In MS MARCO passage v1, 1 $\flops$ amounts to 8.8M posting entries to process for each query.}, 
\begin{equation}
    \operatorname{QD-FLOPs} = \mathbb{E}_{\bQ, \bD}\left( \sum_t \mathbbm{1}(\bw^\bQ_t>0) \times \mathbbm{1}(\bw^\bD_t>0) \right)
\end{equation}
where $\bone(\bw_t^\bQ>0)$ is 1 if the token has a non zero weight in the query $\bQ$ and 0 otherwise (same for $\bD$). We estimate $ \flops$ using a large number of queries and documents. In addition, we also quantify the efficiency by measuring the average number of non-zero components in the documents' representation (we denote it as \emph{Avg. Doc Len} in later sections), to approximate the amount of storage needed for the inverted index.

Our experiments involve the comparison of many different results depending on the architectural and hyperparameter choices. To better compare models in terms of trade-off between effectiveness and efficiency, we therefore designed a new metric, denoted $\EEscore$ (for Efficiency-Effectiveness). Formally, $\EEscore$ is defined as:
\begin{equation}
        \EEscore
        = \mrr - \mu_1 \flops - \mu_2 \softplus_\beta\left(\left(\flops - \tau \right)\right) 
\label{eq:eescore}
\end{equation}
where we rely on softplus~\citep{glorot2011deep}\footnote{We use its Pytorch variant $\softplus_\beta(x) = \frac{1}{\beta}\log(1 + \exp(x))$ that introduces a smoothing parameter~\cite{shamir2022reproducibility}.}, and where $\mu_1$ (before the $\flops$ threshold $\tau$) and $\mu_1+\mu_2$ (after the threshold $\tau$) control the efficiency-effectiveness tradeoff. To illustrate $\EEscore$ let us consider the case where $\beta$, which controls smoothness (in practice, we use $\beta=2$), is large, i.e.,  $\softplus(x)\approx \max(\cdot,x)$. 
In this case, when $\flops=0$, $\EEscore$ is 1 (upper bound) when $\mrr=1$ and 0 when $\mrr=0$. More generally, $\EEscore$ defines the trade-off between $\mrr$ and $\flops$. Let us illustrate this with $\mu_1=0.01$ and $\mu_2=0.09$, which are the values used in this paper. Before the $\flops$ threshold $\tau=5$, $\EEscore$ translates that we are willing to accept a drop of 0.01 $\mrr$ in exchange of an increase in $0.01\times\mu_1^{-1}=1$ $\flops$ , and after $\tau$, of $0.01\times(\mu_1+\mu_2)^{-1}=0.1$ FLOPS – allowing us to define a ``soft'' $\flops$ threshold over which we consider the model to be too costly. 
Although the values we chose ($\tau, \mu_1$, $\mu_2$ and $\beta$) are arbitrary, we think they offer a concise and sound metric for comparing the various models we experimented with.
Finally, in tables and \autoref{fig:pareto_frontier}, we report $\Delta\EEscore$, which is the difference in $\EEscore$ between the evaluated models and BM25. 

\subsection{Premilinary Study on SAE Design for SAE-SPLADE}

In this section, we conduct preliminary experiments to select the best SAE variant (i.e., choosing between TopK SAE and its alternatives) and the SAE training method to be used in the paper.
Across variants, we set the number of SAE latents to $M=2^{16}$ (roughly twice the size of the BERT vocabulary to account for potential dead latents) and use the last hidden state of the base PLM as input (this choice is validated later). We train SAE with $k_{\operatorname{SAE}}=8$ and fine-tune SAE-SPLADE with $k_{\operatorname{SPLADE}}=8$. When $k_{SPLADE}=k_{SAE}$, to reduce clutter, we simply use the notation $k$. 

\subsubsection{Comparing design choices for training SAEs}

SAE works have explored various basic training techniques,  some of the most important being SAE input normalization~\cite{lieberum2024gemma,rajamanoharan2024jumping, bricken2023monosemanticity,gao2025scaling} and training data diversity~\cite{muhamed2025decoding} –- we hence tested their impact.
To normalize the input, we pre-compute an average representation $\overline{\bh}$ and an average norm $\mathbf{\sigma}$ over a set of randomly sampled token representations. During both SAE training and SAE-SPLADE fine-tuning, we use  $\overline{\bh}$ and  $\mathbf{\sigma}$  to normalize $\bh$ as: $\frac{\bh - \overline{\bh}}{\sigma}$, and further rescale the original sparse representation $\bw$ for SAE-SPLADE with $\sigma$ to preserve the magnitude of the vector. 
For the training data, following \citet{park2025decoding} which train the SAE with both query and document texts, we evaluate whether reconstructing both the queries and the documents when pre-training the SAE can help SAE-SPLADE.

We observed that neither modification provides a clear benefit, either in effectiveness or efficiency, to our model.  
This shows that while normalizing and using better training data have both been proven beneficial in other contexts, this does not transfer to the IR task. In particular, we observe that normalization induced a slight increase in the number of ``dead'' latents (around 5\% more).

\subsubsection{PLM's optimal hidden layer} \label{sec:layer_ab}

\newcommand{\Thead}{\mathcal{T}}

\begin{table}[!ht]
    \centering
    \caption{The ablation results for backbone models ($k=8$, TopK SAE) at various layers.  }
    \resizebox{\columnwidth}{!}{
    \begin{tabular}{m{1.5cm}|C{0.8cm}C{0.8cm}C{0.8cm}|C{1.3cm}C{1.0cm}|C{0.8cm}|C{1.2cm}C{1.2cm}}
        \Xhline{1pt}
        Layers & \small{MSM} & \small{TREC-DL} & \small{LoTTE} & \small{QD-flops} $\downarrow$ & 
        \small{Avg. D Len}  $\downarrow$ & \small{$\Delta\EEscore$} $\uparrow$ & Anisot.   $\downarrow$  & Base Anisot.   $\downarrow$ \\
        \Xhline{1pt}
        Layer 6 + $\Thead$ & 37.6 & 71.0 & 68.4 & 0.72 & \textbf{86} & 18.7 & 0.64 & 0.13 \\
        \rowcolor{lightgray} Layer 6 & 37.6 & \textbf{72.1} & \textbf{68.8} & \textbf{0.67} & 109 & \textbf{18.8} & \textbf{0.40} & 0.38 \\ 
        Layer 5 & \textbf{37.7} & 72.1 & 68.8 & 0.86 & 118 & 18.7 & 0.61 & 0.61\\ 
        Layer 4 & 37.2 & 71.8 & 68.0 & 1.01 & 121 & 18.0 & 0.73 & 0.57\\ 
        Layer 3 & 36.8 & 70.2 & 67.3 & 1.13 & 114 & 17.5 & 0.69 & 0.49\\
        \Xhline{1pt}
    \end{tabular}
    }
    \label{tab:sae_splade_layers_ab}
\end{table} 

~\citet{skean2025layer} have shown that different layers of the Transformer architecture play different roles in semantic matching. However, this could not be studied for LSR models since they rely on the MLM head, which contains not only a vocabulary projection layer but also a transform layer $\Thead$ (linearity followed by a GeLU non linear unit), trained to filter out irrelevant information for Masked Token Prediction from the last layer's hidden state.


In contrast, for SAE-SPLADE,  we can decide on which layer to apply the SAE and project its hidden states onto the latent space.
More precisely, instead of training the SAE directly on the last hidden states $\bh_i^L$ of the base encoder, we test training it using the hidden states $\bh_i^l$ of layer $l$ and the ``transformed'' last hidden states $\Thead(\bh_i^L)$.

As shown in \autoref{tab:sae_splade_layers_ab},  we first observe that using $\Thead(\bh_i^L)$ instead of  $\bh_i^L$ hurts all metrics but the document sparsity (Avg. Doc Len) -- demonstrating that it is better to drop the MLM head entirely. Second, the performance is not impacted until we reach layer 4, at which point a noticeable drop occurs. Comparing $\EEscore$  further confirms that using layer 6 achieves the best tradeoff. 

Leaving aside the first configuration (layer 6 + $\Thead$), we  observe that reducing the number of layers consistently increases the QD-FLOPs of the trained model, as well (but to a lesser extent) the average document length.
To better understand this effect, we computed the anisotropy~\cite{godey2024anisotropy} of token representations by measuring the average similarity between random pairs of token representations in multiple contexts for each fully trained model from layer $l$. The hypothesis is that the higher the anisotropy, the harder it is for the SAE encoder to distinguish the different concepts well, which is compensated for by an increase in $\flops$ and a decrease in document sparsity. We report in \autoref{tab:sae_splade_layers_ab} the anisotropy at the layer used by the SAE encoder and observe that this is indeed the case.

Interestingly, for  layer $6 +\Thead$, the model has converged towards sparser document representations while having higher flops than for layer 6, while maintaining an anisotropy comparable to that of layer 5. We hypothesize that this is due to the ``cleaning'' effect of $\Thead$, which is used just before projecting onto the vocabulary. By being constrained (implicitly) by the vocabulary, SAE-SPLADE increases the interaction between queries and documents by re-using similar tokens more often (the query size is stable across models), leading to a degraded $\flops$ metric. We also hypothesize that this might be one of the potential origins of the ``wacky weight'' problem in SPLADE~\cite{mackenzie2021wacky}, where some tokens are over-activated, leading to an overall degraded efficiency.

Overall, these experiments justified the choice of the training SAE at the last encoder layer, i.e., layer $6$ in the case of DistillBERT. Whether this holds true for larger models needs to be studied further.

\subsubsection{Impact of the size of the vocabulary}

\citep{jang2021ultra} show that a higher number of latents correlates with better IR performance. We tested the validity of this claim for SAE-SPLADE with three latent vocabulary sizes $M$, namely $2^{15}, $ $2^{16}$, and $2^{17}$. We only observed a slight effectiveness increase on TREC-DL (around 0.3 point when doubling the size of the vocabulary). We also observed similar \emph{dead node} ratios, which shows that the model with higher vocabulary size tends to make better use of the latent space. However, we did not observe a gap between the QD-FLOPs and Avg. Doc Len. Finally, since a larger $M$ incurs a higher encoding latency, we used $M=2^{16}$ in all subsequent experiments.

\subsubsection{Impact of $k$ and Variants of SAE} \label{sec:k_effect}

\begin{table}[!ht]
    \centering
    \caption{The ablation results of SAE various for different $k$. The best metrics among the SAE variants are in bold, while we \underline{underlined} the best metrics sharing the same $k_{\operatorname{SPLADE}}$. We highlight the three model we present in the main result table. 
    }
    \resizebox{\columnwidth}{!}{
    \begin{tabular}{C{1cm}C{1cm}|C{1.1cm}C{1.2cm}C{1.1cm}|C{1.5cm}C{1.5cm}|C{0.8cm}}    
        \Xhline{1pt}   
        $k_{SAE}$ & $k_{SPLADE}$ & \small{MSM} & \small{TREC-DL} & \small{LoTTE} & \small{QD-FLOPs} $\downarrow$ &
        \small{Avg. D Len} $\downarrow$ & \small{$\Delta\EEscore$} $\uparrow$\\ 
        \Xhline{1pt}
        \multicolumn{8}{l}{\emph{TopK SAE}}\\ \hline
        \multicolumn{2}{c|}{2} & \underline{35.2} & \underline{67.6} & \underline{66.0} & \textbf{0.15} & \textbf{37} & \underline{16.9} \\
        \rowcolor{lightgray} \multicolumn{2}{c|}{4} & \underline{37.1} & \underline{71.5} & \underline{68.2} & 0.33 & 66 & \underline{18.6} \\
        \rowcolor{lightgray} \multicolumn{2}{c|}{8} & \underline{37.6} & \underline{72.1} & 68.8 & 0.67 & 109 & \underline{\textbf{18.8}} \\
        \rowcolor{lightgray} \multicolumn{2}{c|}{16} & \underline{\textbf{38.2}} & \underline{72.3} & 69.4 & \underline{1.37} & \underline{191} & \underline{18.7} \\ 
        \multicolumn{2}{c|}{32} & 37.8 & 72.6 & \underline{\textbf{70.6}} & 2.67 & 316 & 16.9 \\
        8 & $M$ & \underline{38.1} & \underline{\textbf{73.5}} & \underline{70.3} & 5.36 & \underline{369} & 9.5 \\ 
        \Xhline{1pt}
        \multicolumn{8}{l}{\emph{Hierarchical TopK SAE}}\\ \hline
        \multirow{6}{*}{32} & 2 & 34.0 & 67.0 & 65.4 & \underline{\textbf{0.13}} & \underline{\textbf{31}} & 15.7 \\ 
        & 4 & 37.0 & 71.2 & 68.1 & \underline{0.30} & 62 & 18.5 \\
        & 8 & 37.5 & 71.2 & \underline{69.2} & \underline{0.64} & 117 & \textbf{18.7}\\
        & 16 & 37.8 & 71.6 & 69.5 & 1.43 & 208 & 18.2\\
        & 32 & \underline{37.9} & 71.7 & 70.0 & 2.71 & 308 & \underline{16.9} \\
        & $M$ & \textbf{38.0} & \textbf{71.9} & \textbf{70.3} & \underline{4.57} & 387 & \underline{13.7}\\
        \Xhline{0.5pt}
        \multirow{3}{*}{128} & 4 & 37.0 & 71.2 & 67.5 & 0.32 & \underline{56} & 18.5 \\ 
        & 8 & 37.3 & 71.8 & 69.0 & 0.72 & \underline{106} & 18.4 \\
        & 16 & 38.0 & 71.9 & 69.6 & 1.52 & 192 & 18.3 \\
        \Xhline{1pt}
        \multicolumn{8}{l}{\emph{Matryoshka TopK SAE}}\\ \hline
        \multirow{6}{*}{32} & 2 & 34.5 & 67.5 & 65.2 & \textbf{0.14} & \textbf{32} & 16.2 \\
        & 4 & 37.0 & 71.5 & 67.8 & 0.32 & 66 & 18.5\\
        & 8 & 37.5 & 72.0 & 68.9 & 0.73 & 124 & \textbf{18.6}\\
        & 16 & 37.7 & 72.2 & \underline{69.6} & 1.50 & 209 & 18.0\\
        & 32 & 37.7 & \underline{72.8} & 69.6 & \underline{2.60} & \underline{298} & 16.9\\
        & $M$ & \textbf{37.7} & \textbf{73.1} & \textbf{70.0} & 4.84 & 408 & 12.3\\
        \Xhline{1pt}
    \end{tabular}
    }
    \label{tab:sae_splade_topk_ab}
\end{table}

TopK SAE has a limitation: each document token vector has at most $k$ active (>0) latent dimensions, which might be sub-optimal for SAE-SPLADE. At the same time, choosing a good value for $k$ is difficult: on the one hand, if $k$ is too small, the model may fail to capture the complex semantics of the input, worsening the IR performance. On the other hand, if $k$ is too large, it may include irrelevant latent dimensions, thereby increasing the latency of the query.  

To find the optimal $k_{\operatorname{SAE}}$ value for SAE-SPLADE, we vary the value of $k_{\operatorname{SAE}}$ over a wide range, from $k_{SAE}=2$ to $32$, to explore potential trade-offs between effectiveness and efficiency. Each SAE-SPLADE is fine-tuned using $k_{SPLADE}=k_{SAE}$ . We also look at the case where  $k_{SPLADE}=M=2^{16}$ with $k_{SAE}=8$ to maximize the potential number of active latents during IR fine-tuning while keeping the number of activated ones low for SAE reconstruction.





\begin{figure}
    \centering
    \includegraphics[width=\columnwidth]{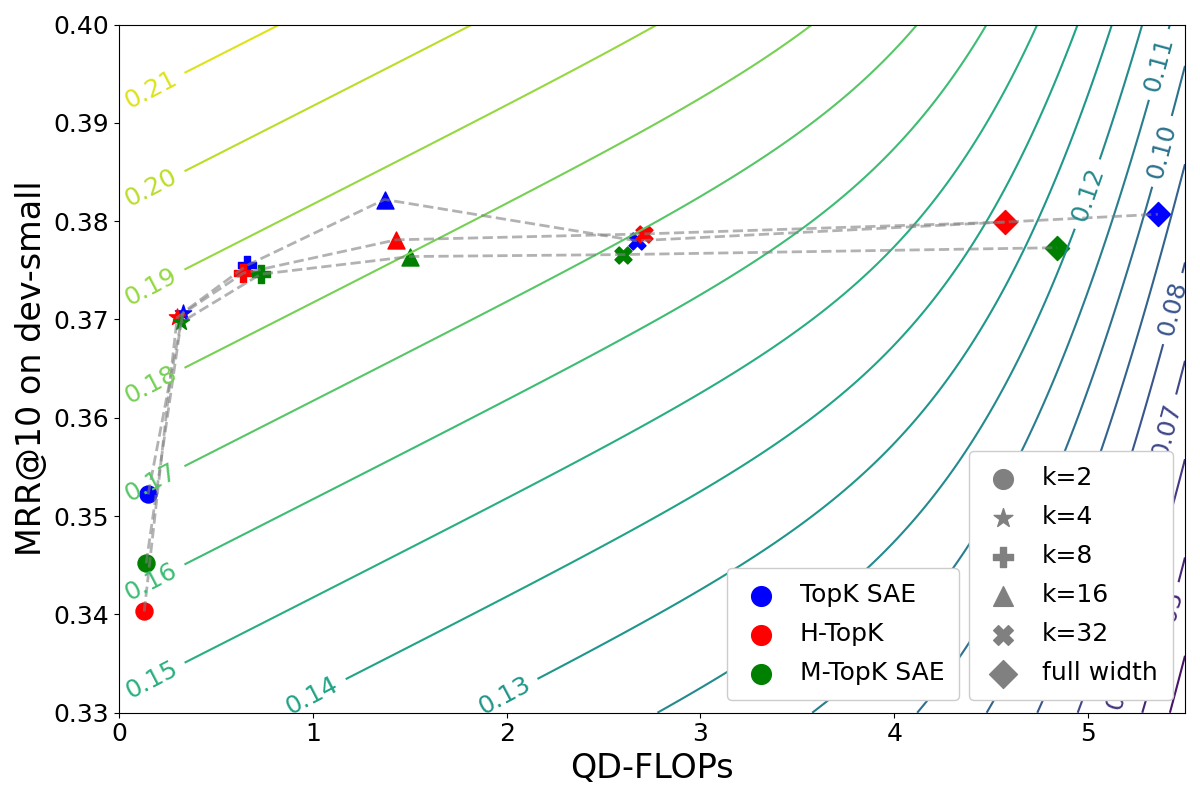}
    \caption{SAE-SPLADE performance depending on the value of $k_{\operatorname{SPLADE}}$ and the TopK SAE variants; contour lines correspond to the $\Delta\EEscore$  (see the definition of the metric in Eq.~\eqref{eq:eescore}). } 
    \label{fig:pareto_frontier}
\end{figure}


Besides the choice of $k$, there are various competing choices for training our encoder. In addition to TopK SAE, we also analyze the results of Hierarchical TopK SAE~\cite{balagansky2025train}, which enforces disentangled latent dimensions across multiple sparsity levels, and Matryoshka TopK SAE~\cite{bussmann2025learning}, which encourages latents to capture semantics across different granularities. The potential advantage of such SAE models is that we can learn the encoder with a higher $k_{SAE}$  before finetuning with a lower $k_{SPLADE}$ since latents are structured. Hence, we initially use $k_{\operatorname{SAE}}=32$  for Hierarchical TopK SAE and Matryoshka TopK SAE\footnote{Following the original work~\citep{bussmann2025learning}, we use $M_i = \left(2048, 6144, 14336, 30720, 65536\right)$, where $M_i$ represents the size of the vocabulary for each nested-SAE model.}.

~\autoref{tab:sae_splade_topk_ab} shows the crucial role of the choice of $k$ in controlling both the document's sparsity (Avg. D Len) and the efficiency of the model (QD-FLOPs).  
Quite naturally, reducing sparsity (i.e., increasing $k_{SAE}$ and $k_{SPLADE}$) consistently improves retrieval effectiveness, highlighting the trade-off between efficiency and performance. 
Secondly, there is no clear advantage in using TopK alternatives: the same $k_{SPLADE}$ value results in a similar performance in terms of effectiveness and efficiency (best $\Delta\EEscore \approx 18.7$ whatever the SAE variant). As Hierarchical TopK had more promising performance, we also trained it with larger $k_{SAE}=128$ to study whether having a larger pool of structured latents could improve SAE-SPLADE performance. We observe that again, there is no clear advantage.
As illustrated in \autoref{fig:pareto_frontier}, and based on our proposed $\Delta\EEscore$, we observe that using $k=4$, $k=8$ or $k=16$ achieve the best trade-off between effectiveness and sparsity for TopK SAE (as they also all have a $\Delta\EEscore > 18$).

To summarize, although the variants of TopK SAE  theoretically improve the structure of the latent space, 
there is no impact on SAE-SPLADE performance. We therefore use the simplest variant, namely TopK SAE, in later sections. We also set $k_{SAE}=8$ in the remainder of this paper as this achieves the best $\Delta\EEscore$.

\subsection{Main Results} \label{sec:main_res}

Based on the preliminary experiments described above, our main SAE-SPLADE model is based on TopK SAE, with $M=2^{16}$ latents, and the SAE encoder uses the DistilBERT layer $\ell=6$ output (before the transform head $\Thead$).  We consider the following baselines for our experiments:
\begin{description}[leftmargin=0.3cm]
    \item[BM25~\citep{Robertson1994OkapiAT}] -- a standard IR baseline;
    
    \item [SPLADE~\citep{formal2021spladev2}] -- the state-of-the-art learned sparse model. We reproduce it with the same training data and objective as SAE-SPLADE for consistency. We further report the result of SPLADEv3 \cite{lassance2024spladev3} based on DistilBERT for comparison;
    
    \item [ColBERTv2~\citep{santhanam2022colbertv2}] -- the state-of-the-art late-interaction model. To ensure fair comparison, we additionally report the results of ColBERTv2 based on DistilBERT from ~\citep{zeng2022curriculum}.
    
    \item [CL-SR~\citep{park2025decoding}] -- an LSR model based on the semantic latents learned with a SAE. Contrary to SAE-SPLADE, CL-SR is based on a single vector dense retrieval model, SimLM~\citep{wang2022simlm}. Consequently, its latents are learned by reconstructing the dense $\texttt{[CLS]}$ token vectors. We only report the performance of the most effective version of the model, as well as the performance of SimLM.
\end{description}

As our baselines span models with very high effectiveness (e.g., SPLADEv3) and  very high efficiency (e.g, BM25), we report results with the 3 different $k$ values ($k=4, 8, 16$ during both the SAE training and the SAE-SPLADE fine-tuning) that achieved the best $\Delta\EEscore$ . Results are reported  \autoref{tab:sae_splade_main}.

\begin{table*}[!ht]
    \centering
    \renewcommand\cellalign{l}
    \caption{The evaluation results for SAE-SPLADE ($k=8$, TopK SAE) and baseline models. We report the metric reflecting the effectiveness (MRR@10 for dev-small, nDCG@10 for TREC-DL mean (DL19 and DL20) and 13 BEIR dataset mean, and Success@5 for 5 LoTTE search test set mean), and efficiency (QD-flops, Avg. D Len), and the metric $\Delta\EEscore$ that reflecting both. The best overall metrics are in \textbf{bold} and the best among ours are \underline{underlined}. 
    Among our SAE-SPLADE models, we use $_\downarrow$ to mark how many sub-dataset we are significantly worse and use $^\uparrow$ to mark how many sub-dataset we are significantly better compare to our SPLADE baseline under the two-tailed Student’s t-test ($p<0.05$).}
    \resizebox{\linewidth}{!}{
    \begin{tabular}{m{2.5cm}m{2cm}|C{1.5cm}C{1.5cm}C{1.5cm}C{1.5cm}|C{1.8cm}C{1.8cm}|C{1cm}}
        \Xhline{1pt}
        \textbf{Model} & \textbf{\small{Encoder}} & \textbf{\small{MSM}} & \textbf{\small{TREC-DL}} & \textbf{\small{LoTTE}} & \textbf{\small{BEIR}} & \textbf{\small{QD-flops}}$\downarrow$ & \textbf{\small{Avg. D Len}} $\downarrow$ & \small{$\Delta\EEscore$} $\uparrow$\\
        \Xhline{1pt}
        \multicolumn{9}{l}{\textit{(Single or Multi Vector) Dense Model}} \\ \hline
        SimLM & BERT\scriptsize{base} & \textbf{41.1} & 70.6 & - & - & - & - & - \\
        ColBERTv2 & DistilBERT & 38.3 & 73.6 & 70.6 & - & - & - & - \\
        ColBERTv2 & BERT\scriptsize{base} & 39.7 & \textbf{75.1} & \textbf{72.0} & \textbf{50.0} & - & - & - \\
        \Xhline{1pt}
        \multicolumn{9}{l}{\textit{Sparse Model}} \\ \hline
        BM25 & - & 18.3 & 49.2 & 51.0 & 43.7 & \textbf{0.13} & \textbf{39} & 0.0 \\
        CL-SR & BERT\scriptsize{base} & 36.8 & 66.0 & - & - & 0.74 & 65 & 17.9 \\
        SPLADEv3 & DistilBERT & 38.7 & 74.8 & 70.3 & \textbf{50.0} & 1.40 & 165 & \textbf{19.1} \\
        \Xhline{1pt}
        \multicolumn{9}{l}{\textit{Ours}} \\ \hline
        SPLADE & DistilBERT & 37.7 & 72.0 & 68.9 & 48.8 & 1.47 & 118 & 18.1 \\
        SAE-SPLADE \scriptsize{$k=4$} & DistilBERT & 37.1\scriptsize{$_{\downarrow 1}$} & 71.5 & 68.2\scriptsize{$_{\downarrow 1}$} & 48.5\scriptsize{$^{\uparrow 1}_{\downarrow 7}$} & \underline{0.33} & \underline{66} & 18.6 \\
        SAE-SPLADE \scriptsize{$k=8$} & DistilBERT & 37.6 & 72.1 & 68.8\scriptsize{$^{\uparrow 1}_{\downarrow 1}$} & 49.2\scriptsize{$^{\uparrow 4}_{\downarrow 2}$} & 0.67 & 109 & \underline{18.8} \\
        SAE-SPLADE \scriptsize{$k=16$} & DistilBERT & \underline{38.2}\scriptsize{$^{\uparrow 1}$} & \underline{72.3} & \underline{69.4}\scriptsize{$^{\uparrow 1}$} & \underline{49.6}\scriptsize{$^{\uparrow 5}_{\downarrow 1}$} & 1.37 & 191 & 18.7 \\
        \Xhline{1pt}
    \end{tabular}
    }
    \label{tab:sae_splade_main}
\end{table*}

SAE-SPLADE ($k=8$) has comparable effectiveness to the SPLADE model we trained, on both ID and OoD tasks, but is slightly better on BEIR (significantly better on 4 BEIR datasets, with $p<0.05$). At the same time, it achieves substantially better efficiency (0.66 vs. 1.47 on $\flops$). Compared with the other sparse models, all our variants largely outperform the BM25 baseline, while our $k=4$ variant achieves a similar index size (Avg. D Len) to CL-SR, but has better efficiency (QD-FLOPs) and improved ID results, especially on the harder TREC-DL datasets.  

Compared to the strong SPLADEv3 baseline (with a $\Delta\EEscore > 19$), our $k=16$ variant achieves almost the same levels of performance: It has very similar $\flops$  (1.37 vs. 1.40), a slightly larger index size, while matching its effectiveness across all datasets we consider, except TREC-DL\footnote{We remind the reader that SPLADEv3 DistilBERT's high performance benefited from many tricks, including curriculum learning~\cite{zeng2022curriculum} and better distillation scores}. 

Our results indicate that SAE-SPLADE underperforms on ID tasks (our $k=16$ variant is on par with DistilBERT-ColBERTv2 on dev-small but lags behind the bigger BERT based ColBERTv2 and SimLM) compared to dense models but is less prone to overfitting (all our variants beat SimLM on the TREC-DL datasets, and SAE-SPLADE $k=16$ achieves very comparable results to the ColBERTv2 models in OoD).
While SAE-SPLADE still falls slightly behind the state-of-the-art late-interaction ColBERTv2 models in effectiveness, late-interaction approaches usually require storing dense token-level embeddings for the entire corpus, resulting in significantly higher disk usage (3 GiB vs. 20 GiB on MS MARCO v1) and retrieval latency compared to our method (<10ms when using BMP index~\cite{mallia2024faster} vs. 58ms when using PLAID~\cite{santhanam2022plaid}).

Taken together, these findings confirm that pre-training an SAE encoder on a large concept vocabulary and then exploiting it to build efficient and effective LSR models is a sound method. The resulting SAE-SPLADE achieves a more favorable effectiveness–efficiency balance than SPLADEv2 while performing on par with the current state-of-the-art LSR approach, SPLADEv3 -- answering both \textbf{RQ1} and \textbf{RQ2}.

\subsection{Ablations}

In this section, we study the hyperparameters that affect the effectiveness and efficiency of SAE-SPLADE models, including the vocabulary pre-conditioning (MLM, SAE, or random), the FLOPs regularization coefficient, and the effect of scaling the backbone encoder's size.

\subsubsection{Vocabulary Pre-Training Effect} \label{sec:vocab_type}

Since masked token prediction in MLM and token representation reconstruction in SAE correspond to different training objectives, we investigate learning  LSR models built on these two types of vocabularies (e.g., SPLADE vs. SAE-SPLADE) by comparing their retrieval effectiveness and efficiency.
To further assess the role of SAE pretraining in the SAE-SPLADE framework, we discard the SAE training stage and fine-tune the model starting from random projections. We apply the same top-$k$ constraints on the token representations during training and inference for all three vocabularies ($k=8$, and with no constraint to match SPLADE). For the models based on the SAE and random vocabulary, we use a vocabulary size of $M=2^{16}$.
\begin{table}[!ht]
    \centering
    \caption{The ablation results of the type of the Vocabulary. ``-'' represent the model too expansive to evaluate. The best results shared by the same vocabulary type are \underline{underlined}, while the best results obtained under the same term-masking strategy ($k=8$ or no mask) are shown in bold.}
    \resizebox{\columnwidth}{!}{
    \begin{tabular}{m{0.6cm}m{0.9cm}|C{1.0cm}C{1.2cm}C{1.0cm}|C{1.3cm}C{1.5cm}|C{0.8cm}}
        \Xhline{1pt}
        \multicolumn{2}{l}{Model} & \small{MSM} & \small{TREC-DL} & \small{LoTTE} & \small{QD-flops} $\downarrow$& \small{Avg. D Len} $\downarrow$& \small{$\Delta\EEscore$}$\uparrow$\\
        \Xhline{1pt}
        \multirow{2}{*}{MLM} & $k=8$ & 37.5 & 71.5 & \underline{\textbf{69.4}} & \underline{0.67} & \underline{\textbf{90}} & \underline{18.7} \\
        & $k=M$ & \underline{37.7} & \underline{72.0} & 68.9 & \textbf{1.47} & \textbf{118} & \textbf{18.1} \\
        \Xhline{0.2pt}
        \multirow{2}{*}{Rand. } & $k=8$ & 36.8 & 71.6 & 68.2 & \underline{1.16} & \underline{134} & 17.5 \\ 
        & $k=M$ & - & \underline{71.8} & \underline{69.4} & 12.88 & 982 & - \\
        \Xhline{1pt}
        \multirow{2}{*}{SAE} & $k=8$ & \textbf{37.6} & \textbf{72.1} & 68.8 & \underline{\textbf{0.66}} & \underline{111} & \underline{\textbf{18.8}} \\ 
        & $k=M$ & \underline{\textbf{38.1}} & \underline{\textbf{73.5}} & \underline{\textbf{70.3}} & 5.36 & 369 & 9.5 \\
        \Xhline{1pt}
    \end{tabular}
    }
    \label{tab:sae_splade_vocab_type}
\end{table}

From \autoref{tab:sae_splade_vocab_type}, we observe that whatever the model, using $k=8$ decreases effectiveness and increases efficiency, as observed earlier. More interestingly, we can draw two conclusions. 
First, the use of a top-$k$ sparsifying approach is generally beneficial for SPLADE, as it substantially increases its efficiency -- decreasing the number of active dimensions for documents by 28, and by 0.8 $\flops$ on average -- and improves its OoD effectiveness (+0.5 for nDCG@10) at the cost of the ID one (-0.5 nDCG@10).
Second, training from a random projection is possible only with a top-$k$ sparsifying approach -- otherwise, the $\flops$ becomes too high (12.88 without, compared to 1.16 with) for practical uses. Still, pre-training with SAE improves both effectiveness and efficiency compared to the random vocabulary.

\subsubsection{FLOPs Regularization Coefficient}

\begin{figure}
    \centering
    \includegraphics[width=\columnwidth]{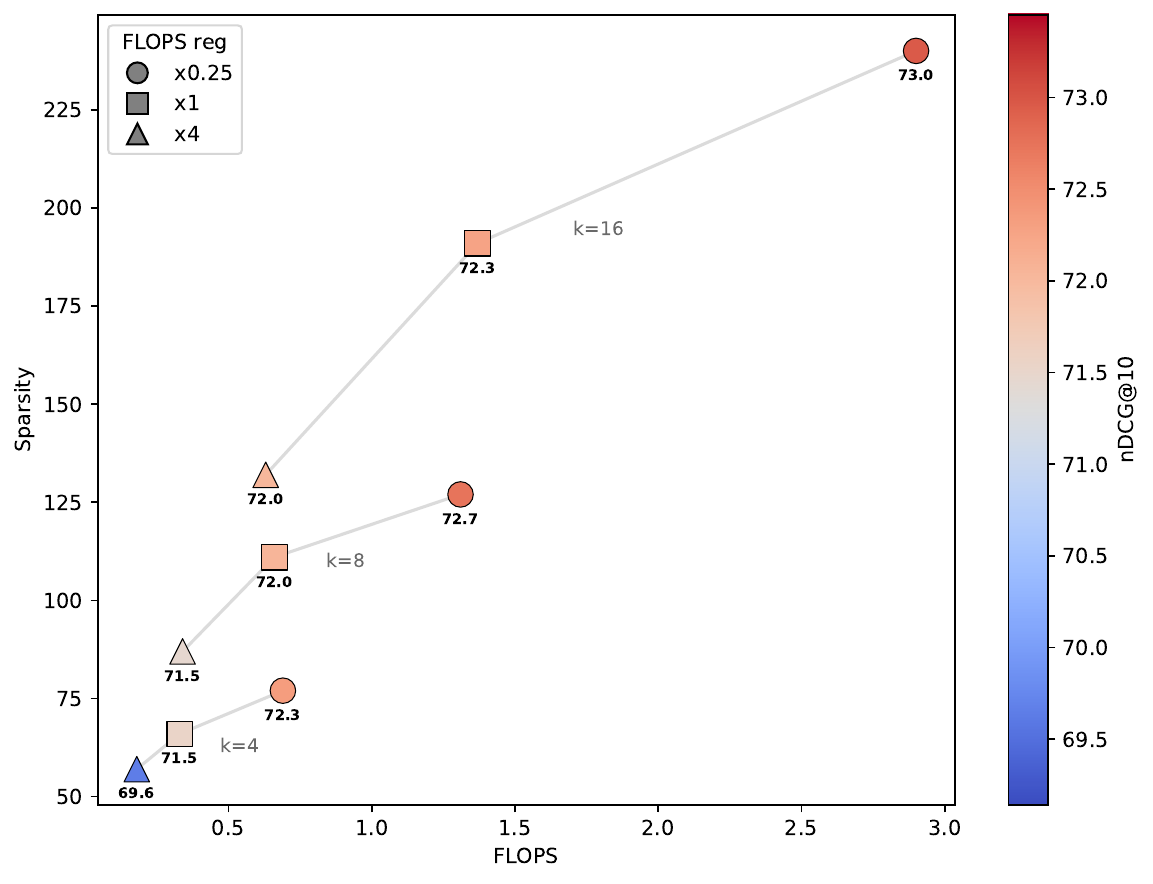}
    \caption{Performance of the SAE-SPLADE model (with TopK SAE) depending on the value of $k = 4, 8, 16$ (in gray, connected by lines) and the FLOPs regularization coefficient (shape).}
    \label{fig:flops_regu_k}
\end{figure}

FLOPs regularization is an effective mechanism for controlling the efficiency-effectiveness tradeoff of LSR models, but its interaction with $k$ is not trivial. We therefore conducted some experiments to investigate their interactions by varying both $k$ and the regularization coefficients $\lambda_{flops-d}$ and $\lambda_{flops-q}$. Results are reported \autoref{fig:flops_regu_k}.

Our results indicate that, as expected, effectiveness increases with larger values of $k$ and decreases with stronger FLOPS regularization, while the opposite trend is observed for efficiency. 
Furthermore, the top-$k$ constraint plays a crucial role in controlling both the sparsity and $\flops$. We find that the model with a small $k$ and a small FLOPs regularization coefficient (0.25x) achieves better sparsity than the models with larger $k$ , even with a high FLOPs regularization coefficient (4x). Meanwhile, as $k$ increases, the impact of the FLOPs regularization on the model becomes more pronounced. 
In general, the TopK constraint is more influential than FLOPS regularization, as we achieve better effectiveness for a given sparsity/$\flops$ budget.
More precisely, under similar QD-FLOPs, the model trained with a smaller $k$ performs better and has a smaller index size (Avg. D Len). This observation suggests that, when controlling the efficiency of the model, reducing $k$ is preferable to increasing the FLOPs regularization coefficient.

\subsubsection{Scaling Effect}

We also compare the performance of SAE-SPLADE ($k=8$) and SPLADE ($k=8$ and full width) with different base model sizes, specifically DistilBERT~\citep{sanh2019distilbert} and BERT~\citep{devlin2019bert}. 
Overall, scaling the base PLM to a larger model yields little difference in ID performance, while providing consistent improvements in OoD settings.
More precisely, for SAE-SPLADE, we observe an average improvement of 0.5 points on LoTTE (68.8 vs. 69.3). In comparison, SPLADE exhibits larger gains when scaling the base model, with improvements exceeding 1.2 points (69.4 vs. 70.2 for $k=8$, and 68.9 vs. 70.1 for $k=M$). Across all cases, the impact on efficiency remains negligible, with differences in QD-FLOPs below 0.05.
These results suggest that scaling to larger base models primarily benefits OoD generalization. 

In conclusion, across all ablation studies, we demonstrate that the SAE-SPLADE is robust across a wide range of training settings and achieves a performance comparable to that of SPLADE in terms of both effectiveness and efficiency. This strengthens our claims on \textbf{RQ1} and \textbf{RQ2}.  
Moreover, we find that applying a token-level top-$k$ constraint during both training and inference boosts the efficiency and effectiveness of SPLADE.


\section{Multilingual Results}

To answer \textbf{RQ3}, regarding whether SAE-based sparse retrieval exhibits better adaptability to multilingual settings than MLM-based models (e.g., SPLADE), we conducted a set of preliminary experiments. 
Specifically, we use a multilingual DistilBERT backbone\footnote{\texttt{distilbert/distilbert-base-multilingual-cased}} and train an SAE-SPLADE-Multilingual model following the same procedure as our main experiments. 
As a baseline, we also train a SPLADE-Multilingual model under identical conditions.
For both models, we apply a token-level top-$k$ constraint with $k=8$.\footnote{We also experimented with SPLADE without a TopK constraint, but the resulting QD-FLOPs were prohibitively high for evaluation.} All other training settings are kept consistent with the monolingual setup described in Section~\ref{sec:train_config}. We also use the automatic translations of MS MARCO, mMARCO~\cite{bonifacio2021mmarco} to train our multilingual SAE and to fine-tune SAE-SPLADE and SPLADE.
Evaluations are performed on the dev-small set of mMARCO and on MIRACL~\cite{zhang2023miracl}, which we use as an OoD benchmark for multilingual retrieval. 
We restrict our training and evaluations to the following languages: \{Arabic (ar), English (en), Spanish (es), French (fr), Japanese (ja), Russian (ru), Chinese (zh)\}, covering both Latin and non-Latin languages. This selection includes languages with overlapping lexical terms (e.g., en/es/fr and ja/zh), as well as languages with totally disjoint vocabularies (e.g., ru and ar), allowing us to assess adaptability across diverse linguistic settings.
\begin{table}[!ht]
    \centering
    \caption{The multilingual performance on mMARCO dev small set. Results of IR performance are in MRR@10. We do not report the ar and ja performance for mColBERT as it is zero-shot in prior work. $\dagger$: significantly better ($p<0.05$) compare to our SPLADE baseline under the two-tailed Student’s t-test. The best across all the baselines are in \textbf{bold} while the best among ours are \underline{underlined}.
    }
    \resizebox{\columnwidth}{!}{
    \begin{tabular}{m{1.9cm}m{0.8cm}|C{0.5cm}C{0.5cm}C{0.5cm}C{0.5cm}C{0.5cm}C{0.5cm}C{0.5cm}|C{0.7cm}}
    \Xhline{1pt}
    \multicolumn{2}{c}{Model} & \small{ar} & \small{es} & \small{fr} & \small{ja} & \small{ru} & \small{zh} & \small{en} & \small{Avg.} \\ 
    \Xhline{1pt}
    \multicolumn{10}{l}{\textit{Prior Works}} \\ \hline
    BM25 & \multirow{2}{*}{\scriptsize{MRR@10}} & 11.1 & 15.8 & 15.5 & 14.1 & 12.4 & 11.6 & 18.3 & 14.1 \\ 
    mColBERT & & - & \textbf{30.1} & \textbf{28.9} & - & \textbf{25.0} & \textbf{24.6} & \textbf{35.2} & - \\
    \Xhline{1pt}
    \multicolumn{10}{l}{\textit{Ours}} \\ \hline
    \multirow{2}{*}{SPLADE \scriptsize{k=8}} & \scriptsize{MRR@10} & \underline{\textbf{19.5}} & 25.6 & 25.2 & \underline{\textbf{23.9}} & 21.1 & 22.9 & \underline{31.4} & 24.2 \\
    & \scriptsize{QDFLOPs} & \textcolor{gray}{5.22} & \textcolor{gray}{2.52} & \textcolor{gray}{2.68} & \textcolor{gray}{4.14} & \textcolor{gray}{3.87} & \textcolor{gray}{3.11} & \textcolor{gray}{2.14} & \textcolor{gray}{3.38} \\ 
    \hline
    \multirow{2}{*}{SAE-SPLADE \scriptsize{k=8}} & \scriptsize{MRR@10} & 18.8 & \underline{26.6}$^\dagger$ & \underline{25.8}$^\dagger$ & 23.6 & \underline{22.0}$^\dagger$ & \underline{23.1} & 30.7 & \underline{\textbf{24.4}} \\
    & \scriptsize{QDFLOPs} & \textcolor{gray}{\textbf{3.94}} & \textcolor{gray}{\textbf{1.66}} & \textcolor{gray}{\textbf{1.88}} & \textcolor{gray}{\textbf{3.20}} & \textcolor{gray}{\textbf{3.02}} & \textcolor{gray}{\textbf{2.09}} & \textcolor{gray}{\textbf{1.63}} & \textcolor{gray}{\textbf{2.49}} \\ 
    \Xhline{1pt}
    \end{tabular}
    }
    \label{tab:ml_mmarco}
\end{table}
\begin{table}[!ht]
    \centering
    \caption{The multilingual performance on MIRACL dev set. Results of IR performance are in nDCG@10. $\dagger$: significantly better ($p<0.05$) compare to our SPLADE baseline under the two-tailed Student’s t-test. The best across all the baselines are in \textbf{bold} while the best among ours are \underline{underlined}.
    }
    \resizebox{\columnwidth}{!}{
    \begin{tabular}{m{1.9cm}m{0.8cm}|C{0.5cm}C{0.5cm}C{0.5cm}C{0.5cm}C{0.5cm}C{0.5cm}|C{0.7cm}}
    \Xhline{1pt}
    \multicolumn{2}{c}{Model} & \small{ar} & \small{es} & \small{fr} & \small{ja} & \small{ru} & \small{zh} & \small{Avg.} \\
    \Xhline{1pt}
    \multicolumn{9}{l}{\textit{Prior Works}} \\ \hline
    BM25 & \multirow{4}{*}{\scriptsize{nDCG@10}} & 39.5 & 7.7 & 11.5 & 31.2 & 25.6 & 17.5 & 22.2 \\
    mContriever & & 52.5 & 41.8 & 31.4 & 42.4 & 39.1 & 41.0 & 41.4 \\
    M3-Sparse~\cite{chen2024bge} & & 67.1 & 38.6 & 35.3 & 56.1 & 44.5 & 36.1 & 46.3 \\
    MILCO~\cite{nguyen2025milco} & & \textbf{80.4} & \textbf{60.9} & \textbf{61.7} & \textbf{77.2} & \textbf{74.6} & \textbf{65.5} & \textbf{70.1} \\
    \Xhline{1pt}
    \multicolumn{9}{l}{\textit{Ours}} \\ \hline
    \multirow{2}{*}{SPLADE  \scriptsize{k=8}} & \scriptsize{nDCG@10} & \underline{59.3} & \underline{45.0} & 38.2 & 49.3 & 47.3 & 43.7 & 47.2 \\
    & \scriptsize{QDFLOPs} & \textcolor{gray}{5.61} & \textcolor{gray}{3.35} & \textcolor{gray}{2.83} & \textcolor{gray}{5.75} & \textcolor{gray}{6.02} & \textcolor{gray}{3.74} & \textcolor{gray}{4.55}\\
    \multirow{2}{*}{SAE-SPLADE  \scriptsize{k=8}} & \scriptsize{nDCG@10} & 58.6 & 44.4 & \underline{38.4} & \underline{50.8$^\dagger$} & \underline{47.4} & \underline{45.9$^\dagger$} & \underline{47.5} \\
    & \scriptsize{QDFLOPs} & \textcolor{gray}{\textbf{4.21}} & \textcolor{gray}{\textbf{2.16}} & \textcolor{gray}{\textbf{1.97}} & \textcolor{gray}{\textbf{4.34}} & \textcolor{gray}{\textbf{4.56}} & \textcolor{gray}{\textbf{2.29}} & \textcolor{gray}{\textbf{3.26}}\\
    \Xhline{1pt}
    \end{tabular}
    }
    \label{tab:ml_miracl}
\end{table}

Results are reported in~\autoref{tab:ml_mmarco} and~\autoref{tab:ml_miracl}.
Across both datasets, SAE-SPLADE achieves slightly better retrieval effectiveness than SPLADE while substantially improving efficiency. These results support our claim that using SAE-derived vocabularies enhances the multilingual adaptability of sparse retrieval models compared to MLM-based vocabularies.
Compared to other baselines, SAE-SPLADE outperforms M3-Sparse~\cite{chen2024bge}, despite the latter relying on a substantially larger backbone model. However, a performance gap remains between our approach and mColBERT~\citep{bonifacio2021mmarco}, a multi-vector dense retrieval model with higher computational cost, as well as MILCO~\cite{nguyen2025milco}, a state-of-the-art multilingual sparse model that employs a larger backbone and more complex training strategies (e.g., cross-lingual alignment) and architectures.\footnote{We note that M3-Sparse and MILCO also leverage the training data from MIRACL for distillation, which is not an OoD evaluation here.}

\begin{table}[!ht]
    \centering
    \caption{The average overlap of vocabulary term over all the languages and the average length of the document representation of our baselines. We report the std using $_\pm$.
    }
    \resizebox{\columnwidth}{!}{
    \begin{tabular}{m{2.5cm}C{2cm}|C{2cm}C{2cm}}
    \Xhline{1pt}
    Model & \small{Voc. Size} & \small{Doc Avg. Len} & \small{Overlap} \\
    \Xhline{1pt}
    SPLADE \scriptsize{$k=8$} & 119547 & 180$_{\pm45}$ & 8.6$_{\pm9.0}$ \\
    SAE-SPLADE \scriptsize{$k=8$} & 65536 & 163$_{\pm43}$ & 11.5$_{\pm8.9}$ \\
    \Xhline{1pt}
    \end{tabular}
    }
    \label{tab:ml_analysis}
\end{table}

We further analyze the representational differences between SAE-SPLADE and SPLADE. Specifically, we randomly sample 2,048 documents from MS MARCO and the corresponding translations in mMARCO, encode them using both models, and compute the number of overlapping activated vocabulary terms or latents with an activation value larger than 0 across all the translations of a document.
The results in~\autoref{tab:ml_analysis} highlight two key observations. 
First, SAE-SPLADE exhibits a higher overlap of activated latents across languages while maintaining stronger sparsity, demonstrating its advantage over SPLADE in learning multilingual representations. This suggests that leveraging SAE-derived latents provides better cross-lingual representation than relying solely on MLM-based lexical vocabularies.
Second, the overlap ratio of 11.5 / 163 indicates that a substantial portion of SAE latents remains language-specific. This finding suggests that learning cross-lingual representations using SAE alone is challenging and that additional cross-lingual adaptation techniques, such as aligning English and multilingual sparse representations used in MILCO~\cite{nguyen2025milco}, could be highly beneficial. We expect that more advanced cross-lingual strategies could be effectively integrated into SAE-SPLADE to further improve its multilingual performance.

In conclusion, all these results and analyzes provide a positive answer to \textbf{RQ3}. 

\section{Qualitative Analysis}

We have seen that it is possible to connect SAE with LSR models and proposed a SAE-SPLADE model that is competitive in terms of effectiveness and efficiency with SPLADE models while being more efficient. The last research question (\textbf{RQ4}) that remains to be studied is understanding the impact of replacing the MLM head with an SAE on SPLADE's behavior.
Our initial hypothesis was that our SAE latents are supposed to encapsulate distinct semantic units, as opposed to tokens, where a single token could sometimes refer to multiple meanings (polysemy), or a single meaning could be covered by multiple tokens (synonymy). 
In this section, we provide a qualitative analysis of the nature of the relationship between SAE-SPLADE latents and SPLADE tokens.

\begin{table}[t]
    \centering
    \caption{First examples (after sorting them by the longest tokens list length) of \emph{synonym} token-latents pairs, grouped by their shared latent.
    }
    \resizebox{\columnwidth}{!}{
    \begin{tabular}{c|c|c}   
        \Xhline{1pt}   
        Latent ID & Synonym tokens list & Concept \\ 
        \Xhline{1pt}
        6470 & ['dish', 'soup', 'pepper', 'ingredients', 'sauce',\ldots, 
        'spices', 'flavors', 'flavour'] & Cuisine \\
        42574 & ['1885', '1880', '1870', '1884', '1881', \ldots 
        '1879', '1875', '1877', '1874'] & 1870-1880s \\
        48320 & ['fry', 'pork', 'stir', 'cooke', 'oven', 'microwave', \ldots, 
        'boil', 'grille', 'barbecue', 'roasted'] & Cooking methods \\
        24849 & ['1902', '1901', '1898', '1899', '1895', '1896',\ldots, 
        '1890s', '1900s'] & Late 19th century \\
        30794 & ['fry', 'pork', 'cooke', 'oven', 'microwave',\ldots, 
        'roast', 'boil', 'grille'] & Cooking methods\\
        16100 & ['moon', 'planet', 'mars', 'orbit', 'planets', 'lunar', 'astronomy',\ldots,
        'astronaut'] & Astronomy \\ 
        9610 & ['breath', 'breathing', '\#\#hale', 'respiratory', \ldots, 
        'pneumonia', 'coughing'] & Respiration \\ 
        \Xhline{1pt}
    \end{tabular}
    }
    \label{tab:synonyms}
\end{table}

In particular, we subsample 10,000 documents from MS MARCO and encode them with both SPLADE and SAE-SPLADE (TopK SAE, $k=8$) models. We use the same rule as for multilingual representations, equating the presence of a token/concept to a weight above 0. To extract meaningful signals, we enumerate the number of occurrences of each token and latent across the entire corpus and remove those with fewer than 5 occurrences.
For each remaining token $t$ and latent $l$, we estimate the distribution of the two conditional probabilities $P(l | T=t)$ and $P(t | L=l)$. 

%
Across our set of token-latent pairs, we further removed any pair with at least one conditional probability below 0.1 to ease our analysis and remove potential noise.
Finally, we run a binomial test over both sets of distributions $P(L|T=t)$ and $P(T|L=l)$ to verify whether the observed co-occurrence values for each pair are abnormal or not, keeping all the pairs for which the hypothesis H0 is rejected with at least 95\% confidence.

We heuristically define "synonymy" as a pair token-latent where $P(t|l) \leq 0.4$ and $P(l|t)\geq 0.6$ (i.e., the token corresponds to the semantic unit covered by the latent, but it is not the only one), "polysemy" as a pair where $P(l|t) \leq 0.4$ and $P(t|l)\geq 0.6$ (i.e., the token is ambiguous and multiple latents cover its different meanings), and "identity" as a pair where $P(l|t)\geq 0.6$ and $P(t|l)\geq 0.6$. For the latter, illustrative token-latent pairs involve tokens such as "canada", "york", and "february". We argue that such SAE latents are necessary to maintain the lexical matching property in SAE-SPLADE.
\autoref{tab:synonyms} provides some examples of the synonym tokens covered by a single SAE latent, while \autoref{tab:polysemy} provides examples of polysemantic tokens somehow disambiguated by latents. Overall, we find that although each individual latent in SAE-SPLADE seems to have its own semantic unit (as illustrated in \autoref{tab:synonyms}), there are "semantic" conflicts between latents, as some tend to cover the same semantic meaning (see, for example, the different latents covering "Cooking methods" in \autoref{tab:synonyms} or the different latents covering the same aspect of the token `bank' in \autoref{tab:polysemy}). 
We hypothesize that this can be caused by a couple of factors, the first being our choice of SAE. Given that we used the simpler TopK SAE approach, it is possible that our latent space is not as structured as it could be with another variant. Secondly, as mentioned previously, this could be the consequence of the contradictions between the SAE learning objective and the IR task fine-tuning. Although we cannot draw conclusive answers for \textbf{RQ4}, our previous empirical results still provide strong evidence that their respective distributions are not aligned, which suggests that the two models do not behave the same. We leave this exploration for future work.

\begin{table}[t]
    \centering
    \caption{Illustrative examples of latents associated to different meanings of a single \emph{polysemantic} token and the corresponding MS MARCO passages.
    }
    \resizebox{\columnwidth}{!}{
    \begin{tabular}{c|c|c}   
        \Xhline{1pt}   
        Token & Latent IDs & Passages \\ 
        \Xhline{1pt}
        \multirow{6}{*}{`spring'} & \multirow{3}{*}{5162} & 'Bonita Springs Tourism: Best of Bonita Springs. Bonita Springs, Florida. [\ldots]' \\
        & & '[\ldots] All King size mattresses use Split box springs. [\ldots]' \\
        \cline{3-3}
         & \multirow{3}{*}{48743} & 'Apr Enlightenment may know no weather, but any pilgrimage is best taken in spring. [\ldots]' \\
        & & ' [\ldots] these invertebrates make up most of a grey fox’s diet in the spring. [\ldots] \\
        \hline 
        \multirow{5}{*}{`bank'} & 23304, 44077, & '1 The Bank was founded in 1934 as a privately owned corporation.[\ldots]' \\
        & 5391, 9394 & '1. Check online for your bank's routing number. [\ldots]' \\
        \cline{3-3}
        & \multirow{3}{*}{55773} & 'In the clip, Banks portrays a sassy real estate-obsessed version of herself [\ldots]' \\
        & & '[\ldots] East of Zebulon they provide access to the Outer Banks, US 64 via Rocky Mount, [\ldots]' \\
        & & '[\ldots] hurricanes are a major threat to the Outer Banks. [\ldots]' \\
        \Xhline{1pt}
    \end{tabular}
    }
    \label{tab:polysemy}
    \vspace{-0.7cm}
\end{table}



\section{Limitation and Conclusion}

In this paper, we present SAE-SPLADE, a variant of SPLADE that replaces the token (lexical) vocabulary with a dictionary of SAE-learned latent representations. 
Using these latents and their activation patterns as a retrieval vocabulary, our method overcomes several limitations of SPLADE based on lexical tokens, including limited scalability to multilingual settings and dependence on the Transformer’s fixed vocabulary.
We conducted extensive experiments to evaluate SAE-SPLADE, showing that it achieves retrieval performance comparable to SPLADE while significantly improving efficiency. In addition, comprehensive ablation studies validate our design choices for both the SAE and the SAE-SPLADE framework. Finally, our results demonstrate that applying a token-level top-$k$ constraint is broadly beneficial for learned sparse retrieval models, including SPLADE itself. 
Overall, our results show that SAE pretraining is a valid alternative to MLM pretraining, allowing the use of potentially fewer layers on larger backbones than DistillBERT, although we did not study this aspect.

Despite these promising results, our work has several limitations and opens up directions for future research.
First, although our multilingual experiments demonstrate that SAE-SPLADE outperforms SPLADE, a performance gap remains with respect to state-of-the-art multilingual models such as MILCO~\cite{nguyen2025milco}. Due to a limited computational budget, we did not incorporate more advanced techniques, such as cross-lingual alignment or more complex model architectures, that have been shown to be effective in prior work. In addition, scaling the multilingual backbone to larger model sizes may further improve performance~\cite{chen2024bge}: recent works such as LACONIC~\citep{xu2026laconic} and SPLARE~\citep{formal2026learning} have shown that learned sparse models can be transferred to large causal language models by adapting bi-directional information for relevance modeling. However, these methods typically require substantially larger index sizes. How to effectively balance the trade-off between effectiveness and efficiency in the new era for LSR remains an open question.

Second, SAEs were originally designed to interpret features in LLMs, and their training objectives are not explicitly aligned with IR. 
Although our experiments show that a straightforward integration of these two research directions, SAE and LSR, produces strong results, in Section~\ref{sec:k_effect}, we demonstrate that some well-designed SAE variants might not be beneficial for IR performance. Future work could further advance this line of research by developing SAE variants specifically tailored to IR objectives.

Finally, SPLADE relies primarily on lexical tokens for exact matching and on expansion mechanisms for capturing semantic similarity. A promising direction is to unify these two signals by combining SAE-based concept learning with lexical-based vocabulary, allowing the model to jointly capture semantic concepts and fine-grained lexical matching cues.



\section*{Acknowledgements}

The authors acknowledge the ANR – FRANCE (French National Research Agency) for its financial support of the GUIDANCE project n°ANR-23-IAS1-0003 as well as the Chaire Multi-Modal/LLM ANR Cluster IA ANR-23-IACL-0007. This work was granted access to the
HPC resources of IDRIS under the allocations 2025-A0191016944 and 2025-AD011014315R2 made by GENCI.

The authors acknowledge the peoples of the Woi Wurrung and Boon Wurrung language groups of the eastern Kulin Nation on whose unceded lands ACM SIGIR 2026 was hosted. We pay our respects to their Elders past and present, and extend that respect to all Aboriginal and Torres Strait Islander peoples today and their continuing connection to land, sea, sky, and community.

\bibliographystyle{ACM-Reference-Format}
\balance
\bibliography{biblio.bib}

\end{document}